\documentclass[reprint,prl,superscriptaddress]{revtex4-1}
\usepackage{graphicx}

\begin{document}

\title{Dynamic particle enhancement in limited-view optoacoustic tomography}

\author{X. Lu\'is De\'an-Ben}
\affiliation{Institute for Biological and Medical Imaging (IBMI), Helmholtz Zentrum M\"unchen, Neuherberg, Germany}
\author{Lu Ding}
\affiliation{Institute for Biological and Medical Imaging (IBMI), Helmholtz Zentrum M\"unchen, Neuherberg, Germany}
\author{Daniel Razansky} \email{Corresponding author: dr@tum.de} 
\affiliation{Institute for Biological and Medical Imaging (IBMI), Helmholtz Zentrum M\"unchen, Neuherberg, Germany}
\affiliation{School of Medicine, Technische Universit\"at M\"unchen (TUM), Munich, Germany}

\date{\today}

\begin{abstract}

Limited-view artefacts are commonly present in optoacoustic tomography images, mainly due to practical geometrical and physical constraints imposed by the imaging systems as well as limited light penetration into large optically opaque samples. Herein, a new approach termed dynamic particle-enhanced optoacoustic tomography (DPOT) is proposed for improving image contrast and visibility of optoacoustic images under limited-view scenarios. The method is based on the non-linear combination of a temporal sequence of tomographic reconstructions representing sparsely distributed moving particles. We demonstrate experimental performance by dynamically imaging the flow of suspended microspheres in three dimensions, which shows promise for DPOT applicability in angiographic imaging in living organisms. 

\end{abstract}


\maketitle 


Due to its hybrid nature combining optical excitation with ultrasonic detection, optoacoustic imaging is capable of visualizing optical absorption contrast in deep tissues with diffraction limited ultrasounic resolution. Optoacoustic images are not affected by speckle-grain artefacts present in backscattering-based coherent imaging techniques, such as pulse-echo ultrasonography or optical coherence tomography, which may hamper the ability to resolve small image features and thus deteriorate the overall image quality. For the latter techniques, the speckle pattern is associated with the superposition of partial waves corresponding to randomly distributed sub-resolution scatterers causing phase shifts in the incident wave ranging from 0 to $2\pi$ \cite{goodman2007speckle,szabo2004diagnostic}. In contrast, bi-polar optoacoustic waveforms mainly accentuate the boundaries of absorbers in A-mode signals, where prominent edges are built up by constructive interference \cite{guo2009speckle}. The inner part of the objects appear then invisible when reconstructions are done by stacking the A-mode optoacoustic signals into B-scan images using e.g. the so-called delay-and-sum algorithms \cite{li2006improved,turner2014improved}. Much like in other tomographic imaging modalities, such as x-ray computed tomography, the imaged object must be fully enclosed by tomographic measurement locations in order to be accurately reconstructed \cite{avinash2001principles}. The so-called limited-view effects would then naturally appear in optoacoustic B-scan images or any tomographic reconstructions lacking full angular coverage \cite{buehler2011model}. These effects become particularly prominent when reconstructing images from objects having elongated structures (e.g. blood vessels), oriented such that they predominantly emit pressure waves in directions that are not covered by detection elements.

Visibility of structures affected by limited-view artifacts can be potentially enhanced by artificially creating small optoacoustic sources within those structures. One approach has used a superposition of multiple images acquired with varying speckled illumination patterns for optoacoustic excitation, where the individual speckle grains represented individual sources \cite{gateau2013improving}. Even though this approach has been experimentally demonstrated in phantoms having a controlled speckle grain size, its applicability for imaging real biological tissues remains challenging due to the need to optoacoustically resolve sub-micron speckle grains in the order of the excitation optical wavelength. A different technique consists in locally heating well-confined spots in the imaged tissue using focused ultrasound, thus thermally encoding the optoacoustic sources via the corresponding local variations of the Grueneisen parameter \cite{wang2015ultrasonic}. The multiple images can then be obtained by scanning the focused ultrasound beam in two or three dimensions, which is generally a lengthy process that may further involve hard compromises with respect to the safety thresholds of focused ultrasound. In any case, it is important to notice that the appropriate image enhancement and improved visibility of structures may only be achieved if the images taken at different time instants are combined in a non-linear manner. A linear superposition operation would merely reduce noise in the images without providing direct benefit in terms of reducing the limited-view artifacts.

In this work we developed a new approach termed dynamic particle-enhanced optoacoustic tomography (DPOT) to improve the visibility in limited-view optoacoustic imaging scenarios. It is based on imaging the dynamic distribution of sparsely located microparticles, which emit optoacoustic waves omni-directionally. The size and average distance between the particles is adapted according to the spatial resolution of the imaging system so that the signals from individual particles can be distinguished. The highest possible concentration of the particles fulfilling the latter criterion would naturally result in the best image quality for a given number of combined frames, with the absorbing structures filled with a dense granular speckle-like pattern for each individual frame.

An illustrative example of the typical case under study is shown in Fig. 1a. It corresponds to a two-dimensional optoacoustic imaging system where signals are collected by an arc-shaped detection array covering a finite angle around the imaged object. In this limited-view scenario, any absorbing object elongated along the central axis of the detection array would predominantly emit optoacoustic waves propagating in directions not covered by the detection aperture, leading to inefficient tomographic collection of signals and invisibility of major portions of the object. Conversely, if small absorbing particles are present within the object, they will emit spherical waves measurable at any detection angle, resulting in better visibility under the limited-view conditions. Fig. 1b shows an example of 100 small particles randomly distributed along a mask mimicking a vascular structure. Superposition of 100 random distributions of the 100 particles leads to the image displayed in Fig. 1c, where the shape of all the vessels is clearly distinguishable.

\begin{figure}
	\centering
		\includegraphics[width=0.45\textwidth]{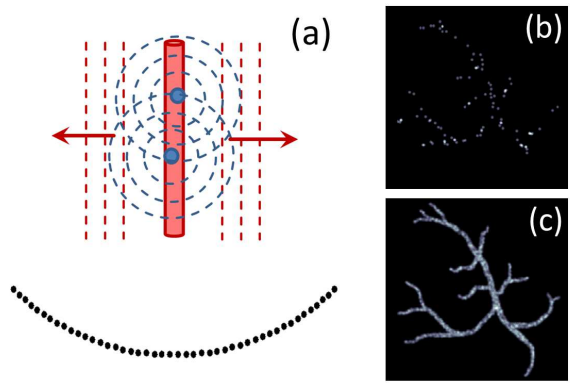}
		\caption{(a) An example of limited-view optoacoustic imaging scenario with an elongated absorbing object oriented along the central axis of an arc-shaped ultrasound detection array (labeled by black dots). The absorbing object would predominantly emit optoacoustic waves in the directions shown by the red dashed lines, leading to inefficient tomographic collection of the generated responses. If small particles (large blue circles) are present within the object, they will emit waves in all directions. (b) An example of random distribution of 100 particles within a mask representing a vascular tree. (c) Superposition of 100 random distributions of 100 particles within the mask.}
	\label{Fig1}
\end{figure}

In practice, the combination of the reconstructed optoacoustic images shall not be done by simple superposition. Even though all the particles may remain visible in the individual images, limited-view tomographic artefacts are usually manifested as negative shadows in the images, canceling out positive contributions in other images. Thus, optimal visibility of structures can only be achieved with a proper non-linear combination of images reconstructed from different random particle distributions. Here we achieved this by simply removing the negative absorption values from optoacoustic images using a non-negative constrained model-based reconstruction algorithm \cite{ding2015efficient}. In this way, the optical absorption in the region of interest for the $i$-th particle distribution, expressed in a vector form $\mathbf{H}_{i}$, is estimated by solving the following least square problem

\begin{equation} \label{eq_recon}
	\mathbf{H}_{i} = \mathrm{argmin}_{\mathbf{H}\geq{}0}\left\|\mathbf{p}_{\mathrm{m}}-\mathbf{A}\mathbf{H}\right\|^2,
\end{equation}

where $\mathbf{p}_{\mathrm{m}}$ is a vector representing the measured signals and $\mathbf{A}$ is the model-matrix corresponding to the discretized linear optoacoustic forward model. The non-linearity in the reconstruction is introduced with the non-negative constraint $\mathbf{H}\geq{}0$. The final image $\mathbf{H}$ of the object is then obtained by simple superposition of the non-negative constrained images from multiple particle distributions, i.e.,

\begin{equation} \label{eq_sum}
	\mathbf{H} = \sum_{i}\mathbf{H}_{i},
\end{equation}

The performance of the suggested method was first tested in a numerical simulation corresponding to the acquisition geometry shown in Fig. 1a. Specifically, 91 equally-spaced measuring positions distributed along a $90^{\circ}$ arc with 40 mm radius were considered. The optical absorption coefficient distribution within the individual 150 $\mu$m radius particles was assumed to be parabolic, for which analytical signals can be analytically calculated \cite{rosenthal2010fast}. Optoacoustic signals corresponding to 100 different random distributions of 100 particles were calculated. Prior to performing the reconstruction according to Eq. 1, the simulated signals were band-pass filtered between 0.25 and 5 MHz in order to further simulate the limited detection bandwidth of common ultrasound detectors. Figs. 2a-c show exemplary images reconstructed from three different random particle distributions. The image obtained by combining all the 100 reconstructions is shown in Fig. 2d. For comparison, the same vascular mask was represented instead by a uniform optical absorption distribution. Image reconstruction was similarly performed with Eq. 1 after downsampling and filtering the signals obtained with the forward model represented by matrix $\mathbf{A}$. The resulting image is displayed in Fig. 2e, showing reduced visibility of vertically-oriented structures as compared to the image enhanced with DPOT.

\begin{figure}
	\centering
		\includegraphics[width=0.45\textwidth]{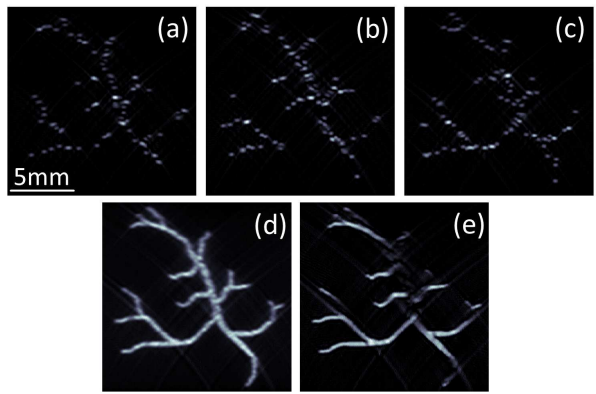}
		\caption{Numerical simulation of model-based optoacoustic image reconstructions in a limited-view detection scenario. (a-c) Non-negative constrained reconstructions obtained from three different random distributions of microparticle absorbers within a vascular structure. (d) Superposition of images obtained from 100 different distributions of the particles. (e) Non-negative constrained reconstruction rendered for the same vascular mask represented by a uniform optical absorption distribution.}
	\label{Fig2}
\end{figure}

The performance of DPOT was experimentally tested using our recently developed real-time three-dimensional optoacoustic imaging platform, which allows for three-dimensional image acquisition at 100 volumes per second rates \cite{dean2013portable,dean2015high}. The imaging system consists of a spherical array of piezocomposite elements covering a solid angle of 90 around the imaged object. The spherical detection aperture has a radius of 40 mm and contains 256 individual detection elements, which are simultaneously sampled at 40 megasamples per second by a custom-made digital acquisition system. While this three-dimensional tomographic configuration substantially improves optoacoustic imaging performance with respect to systems making use of linear or planar arrays \cite{dean2013functional}, limited-view effects still affect the images due to the 90 coverage of the spherical array geometry.

\begin{figure}
	\centering
		\includegraphics[width=0.5\textwidth]{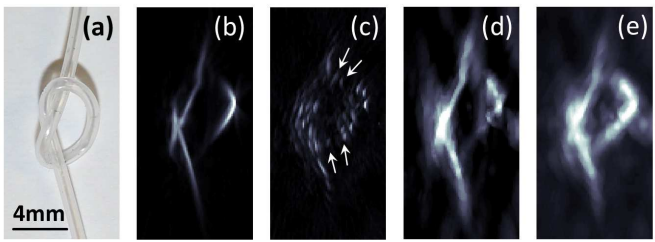}
		\caption{Experimental validation of DPOT using tubing injection of suspended microspheres. (a) Photograph of the tubing used in the experiments. In the experiment, the right part of the tubing in the picture was oriented towards the spherical detection array. (b) Non-negative constrained model-based optoacoustic reconstructions of the tubing filled with ink (maximum intensity projections of the 3D reconstructions are shown). (c) Same reconstruction as in (b) when the suspended 50 $\mu$m diameter absorbing microspheres were used instead. (d) and (e) Corresponding images obtained by superposition of multiple reconstructions for different particle distributions with the unconstrained and non-negative constrained inversions, respectively.}
	\label{Fig3}
\end{figure}

In our experiments, we relied on the powerful capacity of the imaging system for high-speed visualization of flowing particles in 3D. The experimental phantom consisted of a polyethylene tubing (0.6 and 1.1 mm inner and outer diameters, respectively) tangled in a knot and placed in the field of view of the spherical detection array as shown in Fig. 3a. A 10 mm diameter beam from a pulsed laser (Innolas Laser GmbH, Kraling, Germany), tuned for a wavelength of 720 nm and 10Hz pulse repetition frequency, was employed to uniformly illuminate the phantom. Fig. 3b shows a lateral maximum intensity projection (MIP) of the reconstructed three-dimensional optoacoustic image acquired from the tubing filled with black India ink (Higgins, Chartpak, Inc., optical density 5). As can be clearly seen, the lateral sides of the knot are invisible in this image. Furthermore, the image fails to convey the actual thickness of the tubing as it mainly accentuates its boundaries, presumably due to the limited bandwidth of the ultrasound detection array and its limited tomographic coverage. Fig. 3c shows the same MIP for a representative random distribution of 50 $\mu$m diameter polyethylene microspheres (Cospheric BKPMS 45-53) suspended in ethanol. Here multiple particles start appearing in the invisible areas of the knot. Figs. 3d and 3e show the superposition of 80 different microsphere distributions being reconstructed with an unconstrained and non-negative constrained model-based inversion, respectively. A median filter over 5x5x5 voxels was applied to the resulting images in both cases. In the unconstrained case, the linear superposition of images results in reduced visibility of the same areas as in Fig. 3b. In contrast, visibility of the knot is significantly enhanced by superimposing the images reconstructed with non-negative constrained inversion (Fig. 3e), which also better reflects the actual diameter of the tubing. It yet appears that different areas of the knot are not equally visible, which can be presumably attributed to the acoustic distortion in the tubing, agglomeration of spheres or heterogeneous flow. A movie with three-dimensional rotating views for both cases (Movie 1) is available in the online version of the journal for a better qualitative comparison of the images obtained with the ink-filled tubing versus DPOT.

The presented results open up new possibilities for optoacoustic imaging of areas not fully accessible with $360^{\circ}$ tomographic angular coverage. This is of vast importance for clinical translation of optoacoustic imaging, where measurements can only be performed with very limited tomographic coverage of the imaged area \cite{pang2015three,fehm2015volumetric}. In vivo imaging with full angular coverage is often hampered by additional practical factors, such as the mechanical and geometrical constrains or the need to immerse the living organism into a coupling medium \cite{razansky2011volumetric,dean2015non}. Yet, efficient performance of the suggested method in real in-vivo imaging scenarios would imply the use of properly-designed biocompatible particles with sufficiently strong optoacoustic response to be detectable individually. Such particles may serve additional purposes, e.g. provide an efficient optoacoustic feedback to control light intensity distribution through scattering samples \cite{dean2015light}. The signals of individual absorbers can also be used to estimate the flow velocity with Doppler optoacoustic methods \cite{fang2007photoacoustic,yao2010vivo,brunker2012pulsed}, which can also be done with the methodology described here when tracking individual particles in a sequence of time-lapse images. Furthermore, localization of individual particles may further enhance the resolution in optoacoustic tomographic imaging \cite{errico2015ultrafast}.

Future steps must be directed towards improving the spatio-temporal resolution of the optoacoustic imaging system used for DPOT. On the one hand, the effective temporal resolution of the system is correspondingly reduced by the number of images that need to be combined. On the other hand, good spatial resolution is essential to efficiently image dense distributions of small particles, thus the development of a higher resolution imaging system is an important next step. The sensitivity of the optoacoustic system is also an important issue to consider along with the development of efficient particles for DPOT. Indeed, the number of acquisitions required for a good performance of the method is further determined by the signal-to-noise ratio (SNR) of the individual images. The sensitivity can be potentially increased by unmixing spectrally-distinct particles using multispectral image acquisitions \cite{tzoumas2014effects} or photoswitchable substances generating specific temporal profiles \cite{stiel2015high}, for which a very short delay between laser pulses would be necessary in order to avoid motion artefacts \cite{dean2014functional}. Ideally, a system with sufficient resolution and sensitivity to distinguish individual cells would be desirable, as it can potentially enable translating the DPOT method into a setting with only endogenous tissue contrast present.

In conclusion, dynamic particle-enhanced optoacoustic tomography (DPOT) allows improving the visibility of structures affected by limited-view tomographic acquisition artefacts. It thus holds promise for improving accuracy of deep-tissue angiographic imaging provided that the system has sufficient sensitivity to detect signals from individual particles. As limited view effects are unavoidable in most realistic optoacoustic imaging scenarios, DSOT is expected to play an important role in improving the optoacoustic image quality.

This project was supported in part by the European Research Council through the grant agreement ERC-2010-StG-260991.


\end{document}